\documentclass[aps,prd,onecolumn,groupedaddress,showpacs,nofootinbib,amssymb
]{revtex4}
\usepackage[dvips]{graphicx}
\usepackage{amssymb}
\usepackage{amsmath}
\usepackage{graphicx,color}
\usepackage{amsfonts}
\usepackage{bm}

\begin{document}

\tolerance=5000

\title{Dissimilar Donuts in the Sky? Effects of a Pressure Singularity on the Circular Photon Orbits and Shadow of a Cosmological Black Hole}
\author{S.D. Odintsov,$^{1,2,3}$\,\thanks{odintsov@ieec.uab.es}
V.K. Oikonomou,$^{4}$\,\thanks{v.k.oikonomou1979@gmail.com}}
\affiliation{$^{1)}$ ICREA, Passeig Luis Companys, 23, 08010 Barcelona, Spain\\
$^{2)}$ Institute of Space Sciences (ICE,CSIC) C. Can Magrans s/n,
08193 Barcelona, Spain\\
$^{3)}$ Institute of Space Sciences of Catalonia (IEEC),
Barcelona, Spain\\
$^{4)}$ Department of Physics, Aristotle University of
Thessaloniki, Thessaloniki 54124, Greece}

\begin{abstract}
The black hole observations obtained so far indicate one thing:
similar ``donuts'' exist in the sky. But what if some of the
observed black hole shadows that will obtained in the future are
different from the others? In this work the aim is to show that a
difference in the shadow of some observed black holes in the
future, might explain the $H_0$-tension problem. In this letter we
investigate the possible effects of a pressure cosmological
singularity on the circular photon orbits and the shadow of
galactic supermassive black holes at cosmological redshifts. Since
the pressure singularity is a global event in the Universe, the
effects of the pressure singularity will be imposed on
supermassive black holes at a specific redshift. As we show, the
pressure singularity affects the circular photon orbits around
cosmological black holes described by the McVittie metric, and
specifically, for some time before the time instance that the
singularity occurs, the photon orbits do not exist. We discuss the
possible effects of the absence of circular photon orbits on the
shadow of these black holes. Our idea indicates that if a pressure
singularity occurred in the near past, then this could have a
direct imprint on the shadow of supermassive galactic black holes
at the redshift corresponding to the time instance that the
singularity occurred in the past. Thus, if a sample of shadows is
observed in the future for redshifts $z\leq 0.01$, and for a
specific redshift differences are found in the shadows, this could
be an indication that a pressure singularity occurred, and this
global event might resolve the $H_0$-tension as discussed in
previous work. However, the observation of several shadows at
redshifts $z\leq 0.01$ is rather a far future task.
\end{abstract}


\maketitle

\section{Introduction}

Among other mysteries of contemporary research in astronomy and
astrophysics, the $H_0$-tension is the oldest and longstanding at
least for three decades or more. The problem with the value of the
Hubble rate at present day is that large redshift sources, like
the Cosmic Microwave Background (CMB) radiation
\cite{Planck:2018vyg} indicate smaller values compared with small
redshift sources like the Cepheids \cite{Riess:2020fzl}. The
tension can be explained theoretically by early dark energy
\cite{Niedermann:2020dwg,Poulin:2018cxd,Karwal:2016vyq,Oikonomou:2020qah,Nojiri:2019fft}.
Another groundbreaking explanation could be the abrupt change of
physics before $70-150\,$Myrs
\cite{Perivolaropoulos:2021jda,Perivolaropoulos:2021bds,Perivolaropoulos:2022vql}
(see also \cite{Odintsov:2022eqm}) which radically affected the
Cepheid parameters, thus yielding a larger value for the value of
the Hubble rate at present day. Although the $H_0$-tension may be
attributed to the Cepheid calibration
\cite{Mortsell:2021nzg,Perivolaropoulos:2021jda}, such an issue
attracts the attention of cosmologist and astrophysicists, see for
example Refs.
\cite{Dai:2020rfo,He:2020zns,Nakai:2020oit,DiValentino:2020naf,Agrawal:2019dlm,Yang:2018euj,Ye:2020btb,Vagnozzi:2021tjv,
Desmond:2019ygn,OColgain:2018czj,Vagnozzi:2019ezj,
Krishnan:2020obg,Colgain:2019joh,Vagnozzi:2021gjh,Lee:2022cyh,Nojiri:2021dze,Krishnan:2021dyb,Ye:2021iwa,Ye:2022afu,Verde:2019ivm,Marra:2021fvf}
and references therein.

In a recent work \cite{Odintsov:2022eqm} we adopted the
perspective of Refs.
\cite{Perivolaropoulos:2021jda,Perivolaropoulos:2021bds,Perivolaropoulos:2022vql}
that an abrupt physics change before $70-150\,$Myrs might explain
in a transparent way the $H_0$-tension, and we theorized that such
an abrupt and global in the Universe physics change might have
occurred if the Universe experienced a pressure singularity (also
known as Type II or sudden singularity). Such a singularity is not
of the crushing type, and the only physical quantity that is
divergent globally on the spacelike three dimensional hypersurface
that is defined by the time instance at which the singularity
occurs, is the pressure. The rest of the physical observables are
finite, hence this can be viewed as a singularity that the
Universe passes relatively smoothly through it.

In this article we shall examine the qualitative effect of such a
singularity on galactic supermassive black holes, and specifically
on the circular photon orbits around the black hole and the
corresponding shadow. Our proposal is based on the idea that if a
pressure singularity occurred in the past, this could have
potentially observable effects on supermassive black holes
corresponding to redshifts around the time instance that the
singularity occurred, so before $70-150\,$Myrs or for redshifts
$z\simeq 0.01$. The absence of the circular photon orbits could
affect the ring of the shadow, so if scientists in the far future
have available the shadows of a large sample of cosmological black
holes for $z\simeq 0.01$, if the shadows of some black holes at a
specific redshift $z_t$ are different than other black hole
shadows at other redshifts, then this could indicate that a
pressure singularity might have occurred at the specific redshift
$z_t$. The determination of the redshift might also determine the
time at which the pressure singularity occurred.

In order to realize technically the above idea, we shall use the
McVittie metric
\cite{McVittie:1933zz,Faraoni:2007es,Kaloper:2010ec,Lake:2011ni,Nandra:2011ui,Nolan:2014maa,Maciel:2015dsh,Nolan:2017rtj,Perlick:2018iye,Perez:2021etn,Bisnovatyi-Kogan:2018vxl,
Tsupko:2019mfo,Perez:2019cxw,Perlick:2021aok,Nojiri:2020blr}. The
motivation is rather simple, the expansion of the Universe at
cosmological scales should somehow affect the cosmological black
holes. So a more concrete approach for the shadow of the black
hole of cosmological black holes should also take into account the
expansion of the Universe, thus both the gravity of the black hole
and the cosmic expansion at each part of the light trajectory. The
shadow of a black hole is basically a dark spot in the direction
of a black hole in the sky, which is viewable due to the
background of other light sources nearby. The McVittie metric
\cite{McVittie:1933zz,Faraoni:2007es,Kaloper:2010ec,Lake:2011ni,Nandra:2011ui,Nolan:2014maa,Maciel:2015dsh,Nolan:2017rtj,Perlick:2018iye,Perez:2021etn,Bisnovatyi-Kogan:2018vxl,
Tsupko:2019mfo,Perez:2019cxw,Perlick:2021aok,Nojiri:2020blr} is
the most refined description for describing a black hole embedded
in an expanding Friedmann-Robertson-Walker background. For a
mainstream of articles on the McVittie metric, see Refs.
\cite{McVittie:1933zz,Faraoni:2007es,Kaloper:2010ec,Lake:2011ni,Nandra:2011ui,Nolan:2014maa,Maciel:2015dsh,Nolan:2017rtj,Perlick:2018iye,Perez:2021etn,Bisnovatyi-Kogan:2018vxl,
Tsupko:2019mfo,Perez:2019cxw,Perlick:2021aok,Nojiri:2020blr} and
references therein. Although initially it was debatable whether
the McVittie metric describes a black hole \cite{Faraoni:2007es},
it is now widely accepted that indeed it describes a black hole in
an expanding background
\cite{Kaloper:2010ec,Bisnovatyi-Kogan:2018vxl,Tsupko:2019mfo,Perlick:2021aok,Nojiri:2020blr},
although the accretion of matter and radiation is not allowed. The
description of the McVittie solution as a black hole is further
supported by the geodesically incompleteness of the McVittie
metric, due to the existence of a null surface at a finite
distance. Weakly gravitating systems the size of which is small
compared to the comoving Hubble radius are not affected by the
expansion of the Universe in a FRW Universe, however this is not
the case for large scale structures. Indeed, in large scale
structures and cosmological black holes, the effects of the
expansion must be taken into account. Each galactic black hole,
which is a supermassive one, tracks the orbit of the galaxy in the
FRW spacetime, and thus the expansion of the Universe must be
taken into account. The participation of even a strongly bound
local object in the Universes expansion seems to be a general rule
\cite{Faraoni:2007es}. Hence by using the McVittie metric, we
determine whether the condition which allows circular photon
orbits is satisfied or not. As we demonstrate, for a spacetime
with a pressure singularity, the circular photon orbits are not
allowed for some time interval before the singularity, and we
theorize how such a result could affect the shadow of the black
holes being at a specific redshift. We discuss this perspective
and we explain that technically it is hard to verify our proposal
using present day's technology, however in some decades from now,
the resolution techniques will be refined and perhaps our proposal
might be directly investigated experimentally.

\section{Photon Orbits in McVittie's Black Holes and Pressure Cosmological Singularities}

Before discussing the effects of a pressure singularity on the
circular photon orbits in a McVittie black hole, let us recall the
classification of finite-time spacetime cosmological
singularities, following \cite{Nojiri:2005sx}. If the singularity
occurs at the time instance $t=t_s$, we have the following
classification \cite{Nojiri:2005sx}:
\begin{itemize}
\item Type I (``Big Rip'') : A typical crushing type singularity.
As the finite-time singularity is approached at $t \to t_s$, all
the physical quantities that can be defined at the spacelike
hypersurface defined by the time instance $t=t_s$, such as the
total effective pressure $p_\mathrm{eff}$ and energy density
$\rho_\mathrm{eff}$, strongly diverge, including the scale factor
\cite{bigrip}. \item Type II (``sudden''): This is known as the
pressure singularity, firstly studied by Barrow in Refs.
\cite{barrowsudden}, see also \cite{barrowsudden1}. This is the
kind of singularity we shall be interested in this work, in which
case as the singularity is approached, the scale factor and the
energy density is finite, however the pressure diverges. \item
Type III : For this case, as the singularity is approached, the
scale factor is finite, however, both the pressure and the energy
density diverge. \item Type IV : This is a mild singularity
studied in detail in Refs.
\cite{Nojiri:2005sx,Nojiri:2004pf,Barrow:2015ora,Nojiri:2015fra,Odintsov:2015zza,Oikonomou:2015qha,Oikonomou:2015qfh}.
In this case, as the singularity is approached, the scale factor,
the energy density and the pressure are finite, and only the
higher derivatives of the Hubble rate
$\frac{\mathrm{d}^nH}{\mathrm{d}t^n}$ diverge, for $n\geq 2$.
\end{itemize}
In a more transparent way, let us assume that the scale factor has
the following simple form,
\begin{equation}\label{scalefactorini}
a(t)\simeq c(t)+d(t)(t-t_s)^{\eta}\, ,
\end{equation}
where the functions $c(t)$ and $d(t)$ including their higher order
derivatives with respect to the cosmic time are finite at the
cosmic time instance $t=t_s$. Also we assume that
$\eta=\frac{2m}{2n+1}$ with $m$ and $n$ positive integers, in
order to avoid having complex values for the scale factor. The
values of $\eta$ determine the singularity type that may occur at
$t=t_s$. The energy density is affected by the Hubble rate itself
and the pressure by the energy density and the first derivative of
the Hubble rate with respect to the cosmic time. The values of
$\eta$ affect the singularity type in the following way,
\begin{itemize}
\item For $\eta <0$ a Type I singularity occurs, since the scale
factor, the energy density and the pressure are divergent. \item
For $0<\eta<1$ a Type III singularity occurs. \item For $1<\eta<2$
a Type II singularity, or pressure singularity, occurs, since only
the pressure is divergent. \item For $2<\eta$ a Type IV
singularity occurs.
\end{itemize}
Hence, for the pressure singularity one needs $1<\eta<2$, since in
this case at $t=t_s$, only the derivative of the Hubble rate with
respect to the cosmic time diverges. When the Universe goes
through a pressure singularity, it remains geodesically complete,
since the following integral takes finite values for all cosmic
times \cite{Fernandez-Jambrina:2004yjt},
\begin{equation}\label{highercurvsc}
\int_0^{\tau}dt R^{i}_{0j0}(t)\, .
\end{equation}
However, the pressure diverges globally on the spacelike
hypersurface defined by the time instances for which the
singularity occurs.

Now we shall consider the effects of a pressure singularity on the
circular photon orbits around black holes in an expanding
spacetime. We shall discuss the impact of the absence of circular
photon orbits around black holes on the shadow of a black hole,
and also we shall discuss how to verify this observationally on
the shadows of supermassive black holes but in the far future. Our
main assumption will be that a pressure singularity occurred
$70-150\,$Myrs ago, which corresponds to an abrupt physics change.
In the spirit of Ref. \cite{Perivolaropoulos:2021jda}, this might
explain the $H_0$-tension, since an abrupt physics change might
affect directly the Cepheid parameters. So our assumption is that
the pressure singularity indeed occurred before $70-150\,$Myrs, so
at a redshift $z\leq 0.01$.

\begin{figure}
\centering
\includegraphics[width=18pc]{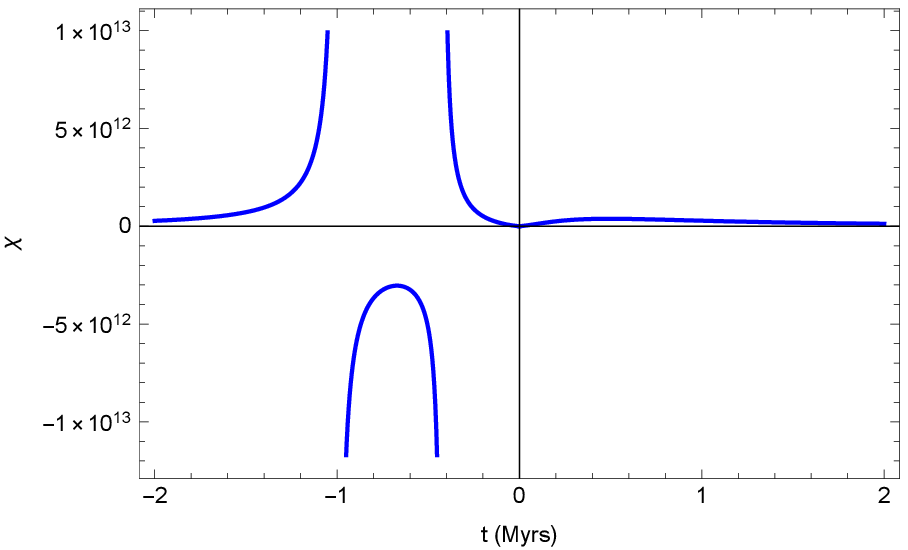}
\includegraphics[width=18pc]{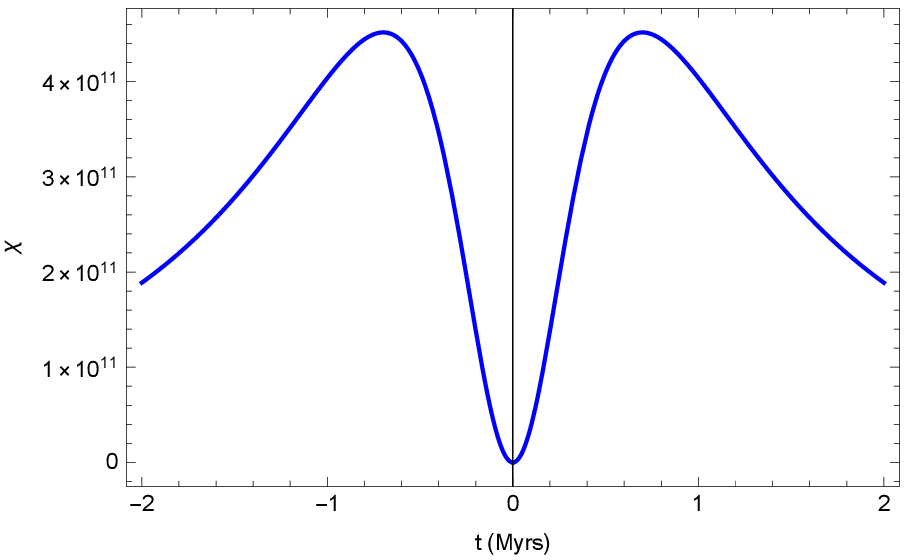}
\caption{ The values of $\chi(R_c,t)$ for $R_c=2.5\,m_0$, for
$\eta=5/2$ (left plot) and for $\eta=2$ (right plot) taking
$c=1.5$ in units of length. The left plot corresponds to the case
for which a pressure singularity occurs at $t=0$. As it is
obvious, for some time before the singularity, the circular photon
orbits do not exist. The qualitative behavior presented does not
change for any radius values in the range $2m_0< R_c\leq
3m_0$.}\label{plot1}
\end{figure}

It is undoubtable at present day that the McVittie metric
describes a black hole in a dynamically expanding
Friedmann-Robertson-Walker (FRW) Universe
\cite{McVittie:1933zz,Faraoni:2007es,Kaloper:2010ec,Lake:2011ni,Nandra:2011ui,Nolan:2014maa,Maciel:2015dsh,Nolan:2017rtj,Perlick:2018iye,Perez:2021etn,Bisnovatyi-Kogan:2018vxl,
Tsupko:2019mfo,Perez:2019cxw,Perlick:2021aok,Nojiri:2020blr}. The
McVittie spacetime metric for a flat FRW background in geometrized
units ($G=c=1$), reads,
\begin{equation}\label{mcvittiemetric}
ds^2=-\left(\frac{1-\frac{m(t)}{2r}}{1+\frac{m(t)}{2r}}\right)^2\cdot
dt^2-\left( 1+\frac{m(t)}{2r}\right)^4 a(t)^2\cdot
\left(dr^2+r^2\cdot (d\theta^2+sin^2\theta d\varphi^2)\right)\, ,
\end{equation}
where the function $m(t)$ is defined as follows,
\begin{equation}\label{mfunction}
m(t)=\frac{m_0}{a(t)}\, ,
\end{equation}
where $m_0$ is the mass of the central body which is embedded in
the expanding spacetime, so basically the mass of the black hole,
and $a(t)$ is the scale factor of the FRW spacetime. When $a=1$
the McVittie metric reduces to the Schwarzschild metric in
isotropic coordinates, while in the limit $m_0\to 0$, the FRW
metric is recovered. Let us now consider the photon orbits in such
a spacetime, and for geodesics paths on the plane
$\theta=\frac{\pi}{2}$, due to the spherically symmetric
spacetime, the conservation of the angular momentum yields,
\begin{equation}\label{conservationofangularmom}
\dot{\phi}=\frac{L}{R^2},\,\,\,\dot{\theta}=0\, ,
\end{equation}
where $R$ is the areal radius coordinate defined as follows,
\begin{equation}\label{arearadiuscoordinate}
R=a(t)r\left(1+\frac{m_0}{2 r a(t)} \right)^2\, .
\end{equation}
The corresponding circular photon orbits geodesics equation reads
\cite{Perez:2021etn},
\begin{equation}\label{circulargeodesciscsphoton}
\frac{L^2}{R^2}=\left(f^2-g^2 \right)\dot{t}^2\, ,
\end{equation}
where the functions $f$ and $g$ are defined as follows
\cite{Perez:2021etn},
\begin{equation}\label{functionsfandg}
f=\sqrt{1-\frac{2
m(t)}{R}},\,\,\,g=R\left(H+\frac{\dot{m}}{m}(f^{-1}-1)\right)\, ,
\end{equation}
with $H$ being the Hubble rate $H=\frac{\dot{a}}{a}$. The quantity
$\chi(R,t)=f^2-g^2=g^{\mu \nu}\nabla_{\mu}r\nabla_{\nu}R$
basically defines the trapped and untrapped spacetime regions of
the spherically symmetric spacetime. The condition for having
circular photon orbits of radius $R_c$ for all cosmic times is
\cite{Perez:2021etn},
\begin{equation}\label{stablecircularorbitscondition}
\chi(R_c,t)=f^2-g^2=1-\frac{2m(t)}{R_c}-R_c^2\left(H(t)+\frac{\dot{m}(t)}{m(t)}\left(\frac{1}{\sqrt{1-\frac{2m(t)}{R_c}}}-1
\right) \right)^2>0\, .
\end{equation}
Obviously in the case that $\chi(t,R_c)<0$, circular photon orbits
cannot exist and as we shall now explain, this will be the case
for cosmic times near a pressure cosmological singularity. In
order to show this explicitly, let us assume that the scale factor
of the Universe is approximately described by,
\begin{equation}
 a(t)=c+ c \vert t \vert^\eta\, , \label{scfactans}
\end{equation}
where $c$ is some arbitrary constant with units $[L]^{-1}$ in
geometrized units and we shall assume that $\eta=\frac{2m}{2n+1}$,
with $n$ and $m$ positive integers. Apparently, for values of
$\eta$ satisfying the condition $1<\eta<2$, a pressure singularity
occurs at the time instance $t=0$. The time instance for which the
singularity occurs is arbitrary, but we chose it to be $t=0$ for
convenience. This time instance $t=0$ can be considered to be any
time instance in the past of our Universe, so for example it can
be before $70-150\, $Myrs. As will now show, depending on the
values of $c$ and $\eta$, the photon orbits with radii $2m<
R_c\leq 3m$ in geometrized units around a black hole in an
expanding Universe might not exist. For the scale factor
(\ref{scfactans}) in Fig. \ref{plot1} we have plotted the quantity
$\chi(R_c,t)$ for $R_c=2.5\,m_0$ and for $\eta=5/2$ (left plot)
and for $\eta=2$ (right plot) taking $c=1.5$ in units of length.
The left plot describes the existence or not of circular photon
orbits for the case that a pressure singularity occurs at $t=0$.
As it is obvious, for a limited time interval before the
singularity, the circular photon orbits do not exist, and now we
shall discuss qualitatively what this result might mean for the
corresponding shadow of the black hole. The qualitative behavior
presented in Fig. \ref{plot1} does not change for any radius
values in the range $2m_0< R_c\leq 3m_0$. What impact would have
the absence of circular photon orbits on the shadow of a
supermassive black hole at cosmological distances? Probably it
would affect the inner part of the shadow, possibly it would
affect the ring of the shadow. Thus the idea that we want to
suggest with this work is simple: by observing a large sample of
supermassive black holes shadows at cosmological distances with
redshifts $z\leq 0.01$, is there any mentionable difference
between the shadows and if yes, at which redshift the shadows are
different? If such a scenario is verified, then this would be an
indication that the Universe in the recent past has experienced a
pressure singularity at the redshift where the singularity
occurred. At the moment, the observation of a large sample of
shadows at cosmological distances is rather technically limited.
Indeed, the observation of the shadow of the M87 supermassive
black hole
\cite{EventHorizonTelescope:2019dse,EventHorizonTelescope:2019ggy}
at $z=0.004283$ is the best outcome that the current technology
can achieve. Thus our proposal cannot be verified at the moment,
since the technical requirements required exceed by far the
current technology. Indeed, in order to capture the shadows of
cosmological galactic supermassive black holes at redshifts $z\leq
0.01$, higher resolutions are required, so more refined VLBI
techniques are required. Current VLBI techniques do not allow to
reach higher resolutions at higher redshifts\footnote{V.K.O. is
thankful to Prof. Luciano Rezzolla for this comment and
perspective.}. Thus our proposal could be investigated
experimentally in the far future, several decades from now. This
includes also the related studies of the M87 and SgrA* which are
too close to us, so their cosmological redshift is not of the
order of $z\sim 0.01$, so not too large in magnitude. Moreover,
the Centaurus A, which is an active galactic nucleus in the nearby
galaxy NGC 5128 with redshift $z\sim 0.00183$, this supermassive
black hole too is relatively chronologically close in the past, so
the effects of a sudden past finite-time singularity would be more
likely spotted on supermassive black holes corresponding to higher
redshifts, of the order $z\sim 0.01$, thus these are future
perspectives of our work, heavily relying on the refined VLBI
techniques.

\section{Future Perspectives}

In this letter we considered the qualitative effects of a pressure
cosmological singularity on the shadow of a galactic supermassive
black hole at cosmological distances. Specifically we considered
the McVittie metric and we investigated what effects would have a
pressure singularity on the circular photon orbits around the
supermassive black hole. As we showed, the circular photon orbits
do not exist for a time interval before the singularity occurred.
This feature can affect directly the shadow of the supermassive
black hole, and specifically we theorized that the pressure
singularity will affect all the supermassive galactic black holes
at the redshift for which the circular photon orbits do not exist.
Thus, if in the far future the shadow of a large sample of black
holes is observed in detail for redshifts $z\leq 0.01$, it is
possible to verify experimentally our conjecture if the
supermassive black holes at a specific redshift show similar
characteristics, which are absent at different redshifts. This can
be a direct indication that at a specific redshift in the past the
Universe experienced a global physics change caused by a pressure
singularity. Thus if such a scenario actually took place before
$70-150\,$Myrs this could solve simultaneously the $H_0$-tension
problem.

Let us now discuss the future perspectives of our theory, since
for the moment it seems impossible to observe a large sample of
shadows with high precision. The effect of expansion of the
Universe on the nearby galactic black holes is tiny, only at
cosmological distances the effect of expansion should be taken
into account. Thus it is compelling to investigate cosmological
black holes at cosmological distances of the order $z\leq 0.01$ in
order to reveal differences between them, to pinpoint the absence
of some characteristic relevant with the shadow of the black hole.
This will reveal the possible effects of the pressure singularity
in the past of our Universe. Precision is required though, to also
take into account the effects of cosmic expansion, the shadow is a
dynamical structure, not static
\cite{Bisnovatyi-Kogan:2018vxl,Tsupko:2019mfo,Perlick:2021aok,Nojiri:2020blr},
however with this paper we aimed to provide a qualitative
description of the whole phenomenon. The whole procedure is a
rather far future task because the current VLBI technique does not
allow to reach different (higher) resolutions at higher redshifts
and also at higher redshifts one must take into account the
influence of the cosmic expansion on the angular diameter of the
black hole shadow \cite{Perlick:2018iye}. This might eventually
cause differences between shadows at low and high redshifts, so
one must also take this into account in order to pinpoint the
effects of a pressure singularity. Moreover a realistic approach
should also take seriously into account the effects of the cosmic
expansion on the surface brightness of light sources
\cite{Perlick:2018iye}, and of course the rotation, but the main
qualitative argument of this work does not change. If some
difference is observed in a sample of different shadows at
redshifts $z\leq 0.01$, this could be due to a pressure
singularity occurring in the near past of our Universe at this
specific redshift.

Although our proposal might be a far future proposal, the current
scientific achievements are encouraging. Indeed, observations of
high redshift supermassive black holes already exist in the
literature \cite{Mortlock:2011va}. Also the shadows of
supermassive black holes at large redshifts will have in general
larger shadows \cite{Bisnovatyi-Kogan:2018vxl}. Also the James
Webb Space Telescope could reveal some properties of supermassive
black holes at large distances. So perhaps in the next decades,
scientists will be able to pinpoint differences in the shadow of
supermassive black holes at redshifts $z\leq 0.01$. The techniques
for obtaining the shadows of black holes are continuously refined
\cite{Younsi:2016azx,Abdujabbarov:2015xqa}, and also theoretical
aspects are further studied \cite{Addazi:2021pty}.

Finally, would be interesting to study the effects of a non-flat
FRW expanding background on the McVittie metric in the case of a
pressure singularity. The same analysis we performed in this paper
should be repeated with non-zero spatial curvature. For
astrophysical black holes, in which case the mass of the black
hole or the radius of the black hole is smaller than the radius of
the curvature, the effects of the curvature are negligible.
However, for cosmological supermassive black holes, the effects of
a non-flat expanding background should be significant. Thus it
would be interesting to repeat the present study in the non-flat
FRW case.

In conclusion, with this work we showed that the two shadows
observed so far, the M87 \cite{EventHorizonTelescope:2019dse} and
the Sagittarius A$^*$ \cite{EventHorizonTelescope:2022xnr} seem
like two identical ``donuts'', the occurrence of a pressure
singularity $70-150\,$Myrs in the past might affect the shadows of
black holes at redshifts $z\leq 0.1$. Thus although it is expected
to see similar ``donuts'' in the sky, if some ``donuts'' are
different, this might be due to a pressure singularity, an effect
which solves simultaneously the $H_0$-tension problem. It is
furthermore interesting to note that one should include the
perspective of pin pointing the modified gravity which may
generate both the McVittie solutions and the past finite-time
singularity, so the most generalized theory is Horndeski theory,
and recently McVittie's solutions were considered in the context
of Horndeski gravity in \cite{Miranda:2022brj}.

\section*{Acknowledgments}

This work was supported by MINECO (Spain), project
PID2019-104397GB-I00 (S.D.O). This work by S.D.O was also
partially supported by the program Unidad de Excelencia Maria de
Maeztu CEX2020-001058-M, Spain.

\end{document}